\documentstyle[prl,aps,twocolumn,amssymb]{revtex}

\def\a{\alpha}
\def\b{\beta}
\def\c{\gamma}
\def\d{\delta}
\def\e{\epsilon}

\def\D{\Delta}

\def\fF{\mathfrak F}
\def\oe{\omega^\epsilon}

\newcommand{\be}{\begin{equation}}
\newcommand{\ee}{\end{equation}}
\newcommand{\bea}{\begin{eqnarray}}
\newcommand{\eea}{\end{eqnarray}}
\newcommand{\beax}{\begin{eqnarray*}}
\newcommand{\eeax}{\end{eqnarray*}}
\newcommand{\op}[1]{\mbox{\sf #1}}
\newcommand{\komm}[2]{\left[#1,#2\right]}

\begin{document}

\draft

\twocolumn[\hsize\textwidth\columnwidth\hsize\csname @twocolumnfalse\endcsname

\title{Time Evolution of Spin Waves}
\author{Michael Moser}
\address{Justgasse 29/12/3/12, A-1210 Vienna, Austria}
\author{Alexander Prets}
\address{Institut f\"ur Theoretische Physik, Universit\"at Wien, 
Boltzmanngasse 5, A-1090 Vienna, Austria} 
\author{Wolfgang L Spitzer}
\address{Department of Mathematics, University of Copenhagen, Universitetsparken 5,
DK-2100 Copenhagen, Denmark}
\date{June, 1999}
\maketitle

\begin{abstract} A rigorous derivation of macroscopic spin-wave equations 
is demonstrated. We introduce a macroscopic mean-field limit and derive the 
so-called Landau-Lifshitz equations for spin waves. We first discuss the 
ferromagnetic Heisenberg model at $T=0$ and finally extend our analysis to 
general spin hamiltonians for the same class of ferromagnetic ground 
states. 
\end{abstract}


]

\narrowtext

\noindent An important problem in statistical physics is to explain the macroscopic 
behavior of matter from first principles, in particular equations of motion of 
macroscopic observables. There is by now an extensive literature on classical 
systems and we want to refer to a standard treatment by Spohn \cite{spohn91}. 
The general idea is that macroscopic equations emerge from microscopic equations 
of motion (say Newton's or Schr\"odinger's) for the constituents of matter by 
performing some kind of scaling limit. There are different types of 
scaling limits whose applicability is to be justified by the physical 
circumstances.

In this paper we study equations of motion for the magnetization in certain
magnetic samples. Long ago Landau and Lifshitz \cite{landau35} derived such 
equations. In case of ferromagnets it says that the change of the magnetization
vector,  $\vec{M}(r,t)$\footnote[1]{$\vec{M}:(r,t) \in {\mathbb R}^d 
\times  {\Bbb R}\mapsto \vec{M}(r,t)\in {\mathbb R}^3$. We set Bohr's magneton 
$\mu_0=\hslash/2$, which puts the constant on the right 
hand side equal to 1.}, is simply given by rotation, i.e.

\be
  \frac{\partial\vec{M}({r},t)}{\partial t} = 
  \vec{M}({r},t)\,\times\,\vec{H}_{\rm eff}({r},t).
\label{l1}
\ee
Here, the angular velocity is equal to an effective magnetic field,
$\vec{H}_{\rm eff}({r},t)$, which originates from the microscopic 
interactions of the spins. If they, say, 
interact only via ferromagnetic exchange interaction \cite{akhiezer68} with 
coupling function $J$, than $\vec{H}_{\rm eff}({r},t)$ is given by 

\be
  \vec{H}_{\rm eff}({r},t) = \int dr'\,J({r},{r}')\,
  \vec{M}(r',t).
\label{l2}
\ee
Expanding $\vec{M}(r',t)$ around $(r,t)$ and considering an isotropic potential,
$J(r,r')=J(|r-r'|)$, $\vec{H}_{\rm eff}$ becomes in lowest
order proportional to the laplacian $\D\vec{M}({r},t)$, and (\ref{l1}) 
simplifies to

\be\label{l3}
  \frac{\partial}{\partial t}\vec{M}(r,t)=\a\,\vec{M}(r,t) \,\times\, \Delta 
  \vec{M}(r,t)
\ee
with
$$
  \a=\frac{1}{6}\int dr\,J(|{r}|)\,r^2.
$$
Usually, (3) is called the Landau-Lifshitz equation. 

We are aware of only two other sources \cite{akhiezer68,kittel}, where
a derivation of equation (\ref{l3}) is demonstrated but, nevertheless, is partly
based on non-rigorous arguments; some of them (e.g. magnon picture), 
however, are extremely important, influential and still a major challenge in 
this field. 
 
We will derive the more general equation (\ref{l1}) by applying what we call 
a macroscopic mean-field limit \cite{spohn91,demasi91,spohn94}. In this context 
we mention that the Vlasov equation for a quantum mean-field 
system has been proven by Narnhofer and Sewell \cite{narnhofer}.
One might think that a direct derivation of equation (\ref{l3}) is 
easier than of (\ref{l1}), but the former does not seem to originate from a 
microscopic model \cite{prets98}. This can be compared with a recent work on 
the derivation of the Cahn-Hilliard equation \cite{giacomin}. We want to 
emphasize that we did not start with a mean-field model in the first place but 
we were forced, both for physical and mathematical reasons,
to do so. We hope that this will become clear in our presentation. 
Admittedly, other approaches may directly lead to equation (\ref{l3}).

Now we provide the set-up for the derivation of equation (\ref{l1}).
Since magnetism is a purely quantum mechanical effect of interacting 
spin systems we clearly have to formulate a quantum mechanical model 
\cite{spitzer96}. For mathematical convenience we consider a bounded sample, 
say the cube $[0,1]^d$. In order to describe the microscopic situation we 
enlarge the cube by a factor $\e^{-1},\e>0$.
Finally, we let $\e\downarrow0$. $K(\e) = [0,\e^{-1}]^d \cap {\mathbb Z}^d$ 
represents the ferromagnet from a microscopic point of view. 
We use the general ferromagnetic Heisenberg-Kac hamiltonian, which is motivated 
by the mean-field structure of (\ref{l2}).

\be
   H^\e=-{\frac{1}{2}}\sum_{x,y\in K(\e)} J_{xy}^\e \, S_x^\a\,S_y^\a,
\label{l4}
\ee
to describe the microscopic exchange interaction energy between spins in 
$K(\e)$. As usual, $H^\e$ acts on the $|K(\e)|$-fold tensor product of the 
space ${\mathbb C}^{2S+1}$. $S$ is some fixed (but arbitrary) half-integer. 
Spin operators $\vec{S_x}=(S_x^1,S_x^2,S_x^3)$ at 
lattice site $x$ (acting on ${\mathbb C}^{2S+1}$) satisfy 
the usual commutation relations
$$
\komm{S_x^\a}{S_y^\b}=i\,\d_{xy}\,\e_{\a\b\c}\,S_y^\c.
$$

We scale the interaction potential like $J^\e_{xy} = {\e}^{+d}J(\e x, \e y)$, 
which can be viewed as a Lebowitz-Penrose approximation \cite{LP} to the true
interaction; the factor $\e^d$ in front of the sum provides us with an extensive
energy function (see also another discussion below). Here, the function $J$ on 
$[0,1]^{2d}$ is assumed to be non-negative but not necessarily symmetric. 

The general idea relating the different levels of description is to define 
the {\it macroscopic} magnetization pro\-file $M^\c(r,t)$ as the $\e$-scaling 
limit of the expectation value of the time dependent {\it microscopic} spin 
operators $S^\c_x(t)$, see (\ref{lim}).

Thus the second main ingredient in our analysis, besides the hamiltonian 
(\ref{l4}), is to specify a class of (initial) states, denoted hereafter by 
$\fF$. It is clear that (\ref{l1}) can only be correct at low temperature.
As is well-known spins are parallel to each other in the (highest weight)
ground state of the (isotropic) ferromagnetic Heisenberg model.
In our states spins will remain almost parallel over short (compared to 
$\e^{-1}$) distances but vary over distances of 
order $\e^{-1}$, i.e. we locally describe $T=0$ equilibrium states as 

\be \label{l8}
  {\fF}=\Big\{(\oe)_{\e>0}:  \oe= \bigotimes_{x\in K(\e)} \oe_x :
  \oe_x(S_x^\c) = M^\c(\e x)\Big\}\nonumber.
\ee
$\oe_x$ is a state on the matrix algebra over ${\mathbb C}^{2S+1}$; in the case 
$S=1/2$ the states $\oe_x$ are uniquely determined by $M^\c$ itself.

We rescale space by $\e$ but we {\em do {not} rescale time.} This particular
space vs time scaling limit fits nicely with Landau's idea of classical spin 
waves of large wave-length, since the microscopic spin-wave length goes as 
${\cal O}(\e^{-1})$. 

In what follows we require 
$M^\c(r,0),\phi(r)$ and $J(r,r')$ to be continuous functions on 
the cube $[0,1]^d$. We now show that at least for 
$0<|t|<C$ there are analytic measures $dM^\c(r,t)$ such that the limits
\be \label{lim}
  \lim_{\e\downarrow0}\e^{d}\sum_{x\in K(\e)}\phi(\e x)\,
  \oe(S^\c_x(t)) = \int_{[0,1]^d} \phi(r) \,dM^\c (r,t)
\ee
for states $\oe\in\fF$ exist. Further, $dM^\c(r,t)$ satisfy the 
Landau-Lifshitz equation (\ref{l1}-\ref{l2}).

Our analysis is based on a series expansion of the time 
evolution of spin operators in finite volume: 

$$
  S^\c_x(t)=
  \sum_{n=0}^\infty\frac{(it)^n}{n!}\,K^n(H^\e)\,S^\c_x,
$$
where $K^n(H^\e)S^\c_x$ means the $n$-th commutator of $H^\e$ with ${S}^\c_x$.
We split this $n$-th commutator into two terms, each of them can be obtained 
recurrently\footnote[2]{This is strictly true only for points $y_1$ such that the 
$l_1$-distance to the boundary of $K(\e)$, $||y_1-\partial K(\e)||_1\ge n$. But,
for fixed $n$, we can always choose $\e$ to be small enough. 
}, 
\bea\label{rec}\lefteqn{ 
  K^n(H^\e)\,S_{y_1}^\c=}
\\
&&(-i)^n\hspace{-1em}{\sum_{y_2,\dots,y_{n+1}\in K(\e)}}^{
  \hspace*{-2em}'}P_n[J^\e](\vec{y}_{n+1})^\c_{\vec{\a}_{n+1}} 
  S^{\vec{\a}_{n+1}}_{\vec{y}_{n+1}}
  +\hat{R}^\e_n.\nonumber
\eea
${\sum_{y_2,\dots,y_{n+1}}}^{\hspace*{-3.5em}'}\hspace*{3.2em}$ means summation 
over all pairwise disjoint variables $y_i\in K(\e)$, i.e.~$y_i\neq y_j$ for all 
$i\neq j$ with $i,j=1,\dots,n+1$, $\vec{y}_{n+1}=(\vec{y}_n,y_{n+1})$, 
$\vec{\a}_{n+1}=(\vec{\a}_n,\a_{n+1})$, and $S^{\vec{\a}_{n}}_{\vec{y}_{n}}=
\prod_{k=1}^{n}S_{y_k}^{\a_k}$. The polynomials $P_n=P_n[J^\e]$ are defined 
in the following way:
\bea \label{l5b}
  P_0(y_1)^\c_{\a_1}&=&\d_{\c\a_1},\nonumber\\*[1em]
  P_1(y_1,y_2)^\c_{\a_1\a_2}&=&\e_{\c\a_1\a_2}\,J_{y_1y_2}^\e,\\*[1em]
  P_n(\vec{y}_{n+1})^\c_{\vec{\a}_{n+1}}&=&P_{n-1}(\vec{y}_n)^\c_{\vec{\b}_n}\,
  \left(T_n^{\a_{n+1}}\right)^{\vec{\b}_n}_{\vec{\a}_n}\,\,n>1
  \nonumber
\eea
with 
$$
T^{\a}_m=\sum_{k=1}^m\underbrace{\op{1}\otimes\dots
\otimes\op{1}}_{k-1\,\mbox{{\scriptsize times}}}\otimes\left(\e^{\a}
\,J_{y_ky_{m+1}}\right)\otimes\underbrace{\op{1}\otimes\dots\otimes
\op{1}}_{m-k\,\mbox{{\scriptsize times}}}.
\label{l9}
$$
$\e^\a$ denotes the matrix of the 
$\e$-tensor, $\e_{123}=1$. The rest term, $\hat{R}_n^\e$, is given by
\bea\label{l5c}
  \hat{R}^\e_n&=&\sum_{k=0}^nK^{n-k}(H^\e)\,R_k^\e,
\\
  R_k^\e&=&
  \frac{1}{4}(-i)^k{\sum_{y_2,\dots,y_k\in K(\e)}}^
  {\hspace*{-1.5em}'}\hspace{1.2em}\sum_{m=1}^k P_k(\vec{y}_k,
  y_m)^\c_{\vec{\a}_k,\a_m}\hat{S}_{y_m}^{\a_m},\nonumber
\eea
with $\hat{S}_{y_m}^{\a_m}=\prod_{m\neq l=1}^kS_{y_l}^{\a_l}$ and 
$R_0^\e=R_1^\e=0$. Formulas (\ref{l5b}-\ref{l5c}) can be proven by induction. 
Since there are more $\e^d$-factors than sums the operator norm 
of the rest term vanishes as $\e\downarrow0$,
\be
\lim_{\e\downarrow0}||\hat{R}_n^\e||=0.
\label{l5d}
\ee
We first show that (for $\c=1,2,3$) the maps

\[\phi\mapsto L^\c(\phi,t):=\lim_{\e\downarrow0}\e^{d}\sum_{x\in K(\e)}
  \phi(\e x)\,\oe(S^\c_x(t))
\]
are continuous linear functionals on the Banach space $C([0,1]^d,\|\cdot
\|_\infty)$ (of continuous functions equiped with the maximum norm) for 
sufficiently small $|t|$ (depending on the interaction function $J$ and on the 
spin $S$), which guarantees the existence of measures $dM^\c(r,t)$. This is 
accomplished by roughly estimating $\e^{d}\sum_{x\in K(\e)}|\phi(\e x)\,
\oe(S^\c_x(t))|$ uniformly in $\e$ by $\|\phi\|_1 \cdot \|M\|_\infty (1-2t\,
\|J\|_\infty\,\|M\|_\infty)^{-1}$ (up to ${\cal O}(\e)$) which holds at least 
for $|t|^{-1}>\|J\|_\infty\,\|M\|_\infty=:C^{-1}$; note that there are $2^nn!$ 
contributing terms. This also shows analyticity for $|t|<C$, so we are allowed 
to interchange the $\e$-limit with the Taylor expansion in time.

By using the formulas (\ref{rec}-\ref{l5c}) for the commutator and the product 
property of the states $\oe$, we can then identify the limit   

\bea\label{sum}\hspace{-2em}\lefteqn{
  \lim_{\e\downarrow0}\e^d\sum_{x\in K(\e)}\phi(\e x)\,
  \frac{\partial^{n+1}}{\partial t^{n+1}}\Big|_{t=0}\,
  \oe(S_{x}^\c(t)) =}
\nonumber\\
&&\int_{[0,1]^d} dr\,\phi(r)\,
  \frac{\partial^{n+1}}{\partial t^{n+1}}\Big|_{t=0}\,{M^\c}({r},t) 
\eea
with the $n$-th derivative of the right hand side of (\ref{l1}).

Note, that we can neglect "surface" points (see second footnote) in the sum in 
(\ref{sum}) close to $\partial K(\e)$, since they are of order $\e^{-d+1}$ and 
do not contribute in the limit. 

We certainly believe that everything is true for all times but we couldn't 
rigorously prove it.


We now want to discuss some generalizations of the Landau-Lifshitz equations 
(1-3). First of all they do appear and secondly, we can fully exploit our 
approach to these general cases. This also gives us insights in the relation 
between symmetries of the microscopic hamiltonian and those of the spin-wave 
equation.


Before changing the spin interaction we mention that we can replace the regular 
(hyper)cubic lattice structure ${\mathbb Z}^d$ itself  
by other crystal structures such as polyatomic Bravais lattices. One could also 
immediately start with an infinitely extended sample instead of the cube (the 
test function $\phi$ as well as the interaction $J$ should then be Schwartz 
functions), or more rigorously by additionally performing a thermodynamic limit.


We have just shown how the term $\vec{M}\times\Delta \vec{M}$ descends from  a 
2-point spin interaction. In general, we might have contributions (in raising
complexity) from an exterior magnetic field, $\vec{M}\times\vec{B}$, 
magnetic-anisotropy energy (due to the presence of an axis of easiest magnetization), 
$(0,0,M^3) \times \vec{M}$, and relativistic corrections 
like $\vec{M}\times(\vec{M}\times\Delta\vec{M})$, which determine the 
time derivative of $\vec{M}$. We now explain how these terms emerge from 
general $N$-point spin interactions. To this purpose we start with a hamiltonian
 
\be \label{npoint}
     H^\e_N = \sum_{m=1}^N \sum_{y_1,\ldots,y_m} (J^\e_m)_{y_1\ldots y_m}^{\a_1
     \ldots\a_m}\,S_{y_1}^{\a_1}\cdots S_{y_m}^{\a_m}
\ee
which includes multiple spin interactions up to a maximal, but otherwise 
arbitrary order $N$. Whereas Akhiezer et al derive for instance $\vec{M}\times 
\vec{M} \times\Delta\vec{M}$ by taking into account higher orders in the 
Holstein-Primakoff expansion \cite{akhiezer68} 
we directly consider general $N$-point spin interactions. 

As before the tensor-valued functions $(J^\e_m)_{y_1\ldots y_m}^{\a_1\ldots\a_m}$
are slowly varying functions w.r.t. spatial coordinates and scale in magnitude 
as $\e^{(m-1)d}$ such that the energy is extensive; we continue to write now 
$(J^\e_m)_{y_1\ldots y_m}^{\a_1\ldots\a_m} = \e^{(m-1)d}\,J_m^{\a_1\ldots\a_m}
(\e y_1,\ldots,\e y_m)$.

Surprisingly, one can calculate the time evolution generated by the more general
hamiltonian in (\ref{npoint}) and analyze the scaling limit applied to states in
$\fF$. Even more, the structure of the spin-wave
equation is the same as in equation (\ref{l1}), only the effective magnetic field
has to be generalized to

\bea\lefteqn{\vec{H}_{\rm eff}(r,t) = J_1^{\a_0}(r) + 
  \sum_{m=1}^{N-1} \int dr'_1 \cdots dr'_{m} \,} 
\\&&   
   J_{m+1}(r,r'_1,\ldots,r'_{m})^{\a_0 \a_1\ldots\a_{m}} 
   M^{\a_1}(r'_1,t)\cdots M^{\a_{m}}(r'_{m},t).\nonumber
\eea


\noindent
The formulas for the time evolution generated by $H_N^\e$ are analogous to 
(\ref{rec}-\ref{l5c}) but more complicated in detail and will be omited here.

We remark that the use of product-states as initial states for the Heisenberg 
model is perfectly justified by its local $T=0$ equilibrium properties, but 
in general cases we consider them just as convenient states, 
which seem (without proof) to be close to experimentally realizable initial
states.

All kinds of Landau-Lifshitz equations which originate from hamiltonian 
dynamics (and therefore conserve the magnetization) can now be 
derived by properly choosing the interaction potentials. 
In order to see how this works we now apply this to the interesting case of 
3-point interactions,

\be \label{3point}
  H^\e_3 = \sum_{x,y,z} (J^\e_3)_{x y z}^{\a\b\c}\, S_{x}^{\a} S_{y}^{\b} 
  S_{z}^{\c},
\ee 
which after all is {\it the} motivation for our generalizations. From the
invariance under rotations w.r.t. spin indices, we choose
$(J^\e_3)_{x y z}^{\a\b\c}=:(J^\e_3)_{x y z}\,\e^{\a\b\c}$. It is fair to ask 
for spatial translation invariance of $J_3$, such that $(J^\e_3)_{x y z}=:
J(\e(y-x),\e(z-x))$. Further, it is reasonable that $J$ has the properties
$J(y,z)=-J(z,y)=J(-y,-z)$. [Any odd function, $f$, provides an example 
by setting $J(y,z)=f(|y|^2-|z|^2)$.] After expanding the
integrand in $\vec{H}_{\rm eff}$ at $r$ we find the desired equation 

\[
  \frac{\partial}{\partial t}\vec{M}(r,t) = \a_{ij}\,\vec{M}(r,t)\times\Big(
  \vec{M}(r,t)\times\partial_i \partial_j \vec{M}(r,t)\Big)
\]
with $\a_{ij} = \int dy dz\, J(y,z)\,y_i z_j$.

\bigskip
{\bf Acknowledgments:}
AP greatly acknowledges the support by the Austrian Science Fund (FWF), project
P10517-NAW and wants to thank B.~Baumgartner for organizing and promoting this 
project. We are also indebted to C.~Maes, H.~Narnhofer and A.~Verbeure.

\end{document}